%% file: main.tex
\documentclass[sigconf]{acmart}

\usepackage{booktabs} 
\usepackage{tikz}
\usetikzlibrary{arrows}
\usepackage{color}

\usepackage{graphicx}

\usepackage{listings}

\definecolor{dkgreen}{rgb}{0,0.6,0}
\definecolor{gray}{rgb}{0.5,0.5,0.5}
\definecolor{mauve}{rgb}{0.58,0,0.82}

\setcopyright{none}

\acmConference[]{White paper}{2018}{Brooklyn, NY}
\copyrightyear{2018}

\begin{document}
\title{Koji: Automating pipelines with mixed-semantics data sources}

\author{Petar Maymounkov}
\affiliation{%
  \institution{Aljabr, Inc.}
}
\email{p@aljabr.io}


\begin{abstract}
We propose a new result-oriented semantic for defining data processing workflows that
manipulate data in different semantic forms (files or services) in a unified manner.
This approach enables users to
define workflows for a vast variety of reproducible data-processing tasks in a simple declarative
manner which focuses on application-level results, while automating
all control-plane considerations (like failure recovery without loss of progress
and computation reuse) behind the scenes.

The uniform treatment of files and services as data
enables easy integration with existing data sources (e.g. data acquisition APIs)
and sinks of data (e.g. database services).
Whereas the focus on containers as transformations
enables reuse of existing data-processing systems.

We describe a declarative configuration mechanism, which
can be viewed as an intermediate representation (IR)
of reproducible data processing pipelines in
the same spirit as, for instance, TensorFlow~\cite{tensorflow} and
ONNX~\cite{onnx} utilize IRs for defining tensor-processing pipelines.
\end{abstract}


\maketitle

\input{intro}
\input{example}
\input{related}
\input{semantics}
\input{arch}
\input{conclusion}
\appendix
\input{spec}

\bibliographystyle{ACM-Reference-Format}
\bibliography{bibliography}

\end{document}

%% file: intro.tex
\section{Introduction}

\subsection{History}

The introduction of MapReduce~\cite{mr} by Google arguably marked the beginning of
programmable large-scale data processing.
MapReduce performs a transformation of one set of large
files (the input) into another (the output). Since the transformation provided by a MapReduce
is a primitive — a many-to-many shuffle, followed by an element-wise map — it
became common practice to chain multiple MapReduce transformations in a pipeline.

The dataflow in such a pipeline is cleanly captured by a directed-acyclic graph (DAG),
whose vertices represent transformations and edges represent files.

In a twist, it became commonplace to query a service (usually a key-value lookup service)
from inside the mapper function. For instance, this technique is used to join two tables by
mapping over one of them and looking up into the other. More recently, Machine Learning systems
have been serving trained models as lookup services, which are used by MapReduce mappers in
a similar fashion.

With this twist, a MapReduce transformation no longer depends just on input files but
also on lookup services (and their transitive dependencies, which are usually other files).
The simple dataflow model mentioned previously no longer applies.

To the best of our knowledge, no dataflow model has been proposed to capture this 
scenario. Yet, to this day, this type of mixed-semantic (file and service) pipelines
represent the most common type of off-line batch-processing workflows.

Due to the lack of a specialized formalism for describing them and a tool for executing them,
they are currently codified in a variety of error-prone ways, which usually
amount to the usage of data-unaware task execution pipelines. We address this gap here.

\subsection{Problem}

We address a class of modern Machine Learning and data-processing pipelines.

Such pipelines transform a set of input files through chains of
transformations, provided by open-source software (OSS) for large-scale computation
(like TensorFlow~\cite{tensorflow},
Apache Beam~\cite{apache_beam}, Apache Spark~\cite{apache_spark}, etc.) or user-specific implementations
(usually provided as executable containers). In short, these
pipelines use mixtures of disparate OSS technologies tied into
a single coherent data flow.

At present, such workflows are frequently built using task-driven pipeline
technologies (like Apache Airflow~\cite{apache_airflow}) which execute tasks in
a given dependency order, but are unaware of the data
passed from one task to the next.
The lack of data flow awareness of current solutions prevents
large data-processing pipelines from benefiting in
caching and reuse of computation, which could provide
significant efficiency gains in these industry cases:

\begin{itemize}
	\item Restarting long-running pipelines after failure and
	continuing from where previous executions left off.
	\item Re-running pipelines with incremental changes,
	during developer iterations.
	\item Running concurrent pipelines which share logic, i.e.
	compute identical data artifacts within their workflow.
\end{itemize}

Furthermore, task-centric technologies make it impractically hard
to integrate data and computation optimizations like:

\begin{itemize}
	\item In-memory storage of intermediate results which are not cache-able, or
	\item Context-specific choices of job scheduling and placement algorithms.
\end{itemize}

\subsection{Solution}

We propose a new pipeline semantic (and describe its system
architecture, which can be realized on top of Kubernetes)
based on a few key design choices:

\begin{itemize}
	\item \textbf{Result-oriented specification:}
	The goal of workflows is to build
	data artifacts. Workflows are represented as dependency graphs over
	artifacts. Input artifacts are provided by the caller. Intermediate
	artifacts are produced through a computation, using prior artifacts.
	Output artifacts are returned to the caller. This view is entirely
	analogous to the way software build systems (like Bazel~\cite{bazel} and UNIX \texttt{make}) define
	software build dependencies.

	\item \textbf{Unified treatment of data and services:}
	We view file artifacts and service artifacts in a unified way as \textit{resources}.
	This allows us to describe complex workflows which mix-and-match
	batch and streaming computations (the latter being a special case of services).
	Furthermore, this enables us to automate service garbage-collection
	and achieve optimal computation reuse (via caching) across the entire pipeline.
	The resource-level unified view of files and services purports to be
	the Goldilocks level of coarse data knowledge, that is needed by a dataflow
	controller to automate all file caching and service control considerations.

	\item \textbf{Type-safe declarative specification:}
	We believe that workflow specification has to be declarative, i.e. representable
	via a typed schema (like e.g. Protocol Buffers). This provides
	full decoupling from implementations, and serves as a reproducible
	assembly language for defining pipelines.

	\item \textbf{Decouple dataflow from transform implementation:}
	We decouple the specification of application logic from the definition of how
	data transforms are performed by underlying backend technologies.
	Application logic comprises the dependency graph between artifacts,
	and the data transform at each node. Data transforms are viewed
	uniformly akin to functions from a library of choices.
	The methods for invoking transformation-backing technologies (like MapReduce, TensorFlow, etc.)
	are implemented separately as driver functions,
	and surfaced as a library of declarative structures that can be used in application logic.

	\item \textbf{Extensible transformations:}
	New types of data transforms (other than container-execution based) can be added easily.
	This is done in two parts. First, a simple driver function
	implements the specifics of calling the underlying technology.
	Second, a new transformation structure is added to the application logic schema.
	This extension mechanism is reserved for transformations that cannot be containerized.

	\item \textbf{Scheduler and storage-independent design:}
	Application logic governs the order in which data computations must occur.
	However, The choice of job schedulers (or placement) algorithms,
	as well as the choice of storage technologies (e.g. disk versus memory volumes),
	are entirely orthogonal to the application's dataflow definition.
	Our architecture enables flexible choice of relevant technology
	on a per-node (scheduling) and per-edge (storage) basis.
	For instance, some intermediate files can be stored in memory volumes,
	instead of disk, to increase efficiency.

\end{itemize}

We aim for this to be a pipeline technology which can perform data transformations
based on any software available, as OSS-for-Linux or as-a-Service. This goal informs
our choice of Kubernetes, as an
underlying infrastructure and cluster management technology for coordinating orchestration of backend execution runtimes on single or multi-tenant physical or (multi)cloud computing resources.

Kubernetes~\cite{kubernetes}, which is becoming the industry-standard cluster OS,
is ubiquitously available on most cloud providers, can be run on-premises, and is generally
provider-agnostic from the users' standpoint. Kubernetes benefits from having mature
integrations to the Docker (and other) container ecosystems, providing seamless out-of-box
access to many data tools, thanks to the rich ecosystem of operator implementations~\cite{k8ops}.

\subsection{How it works}

The user defines a pipeline in a language-agnostic manner.
A pipeline definition describes: (i) input data resources and their sources,
(ii) intermediate resources and the data transformation that produced them
from dependent resources, (iii) output data resources and where they should be delivered.

\begin{itemize}
\item Resources are files (or services) and their format (or protocol) can be optionally
specified to benefit from type-safety checks over the dataflow graph, thanks to
declarations.

\item Input resources can be provided by various standard methods (volume files,
existing cluster services, Amazon S3 buckets, and so on.). Data source types
can be added seamlessly.

\item Intermediate resources are described as files (or services), optionally with a 
specified format (or protocol). Their placement location is not provided by the user,
in order to enable the pipeline controller to make optimal choices in this regard
and to manage caching placements decisions.

\item Output resources can specify the location where they should be delivered,
with standard (and extensible) options similarly to the case for input resources.
\end{itemize}

The transformations at the nodes of the dataflow graph consume a set
of input resources to produce new resources. Transformations are 
exposed to the user with a clean application-level interface. A transformation:

\begin{itemize}
	\item Consumes a set of named input resources (the arguments), which can
	be fulfilled by prior resources in the user's dataflow program.

	\item Produces a set of named output resources. The latter can be referenced (by name)
	by dependent transformations, downstream in the dataflow program.

	\item Accepts a transformation-specific set of parameters. For instance,
	a TensorFlow transformation may require a TensorFlow program (e.g. as a Python or Protocol Buffer file).
\end{itemize}

The user programs (in Python or Go) which build the pipeline dataflow
are used to generate a Protocol Buffer (or YAML) file, which captures
the entire pipeline and is effectively executable and reproducible.

Pipelines can be executed either from the command-line or by sending them
off as a background job to Kubernetes, using the operator pattern (via CRD).

%% file: example.tex
\section{An example}

A typical modern Machine Learning pipeline produces
interdependent files (e.g. data tables) and services (e.g. trained model servers)
at multiple stages of its workflow.

\begin{figure}[h]
\includegraphics[width=6cm]{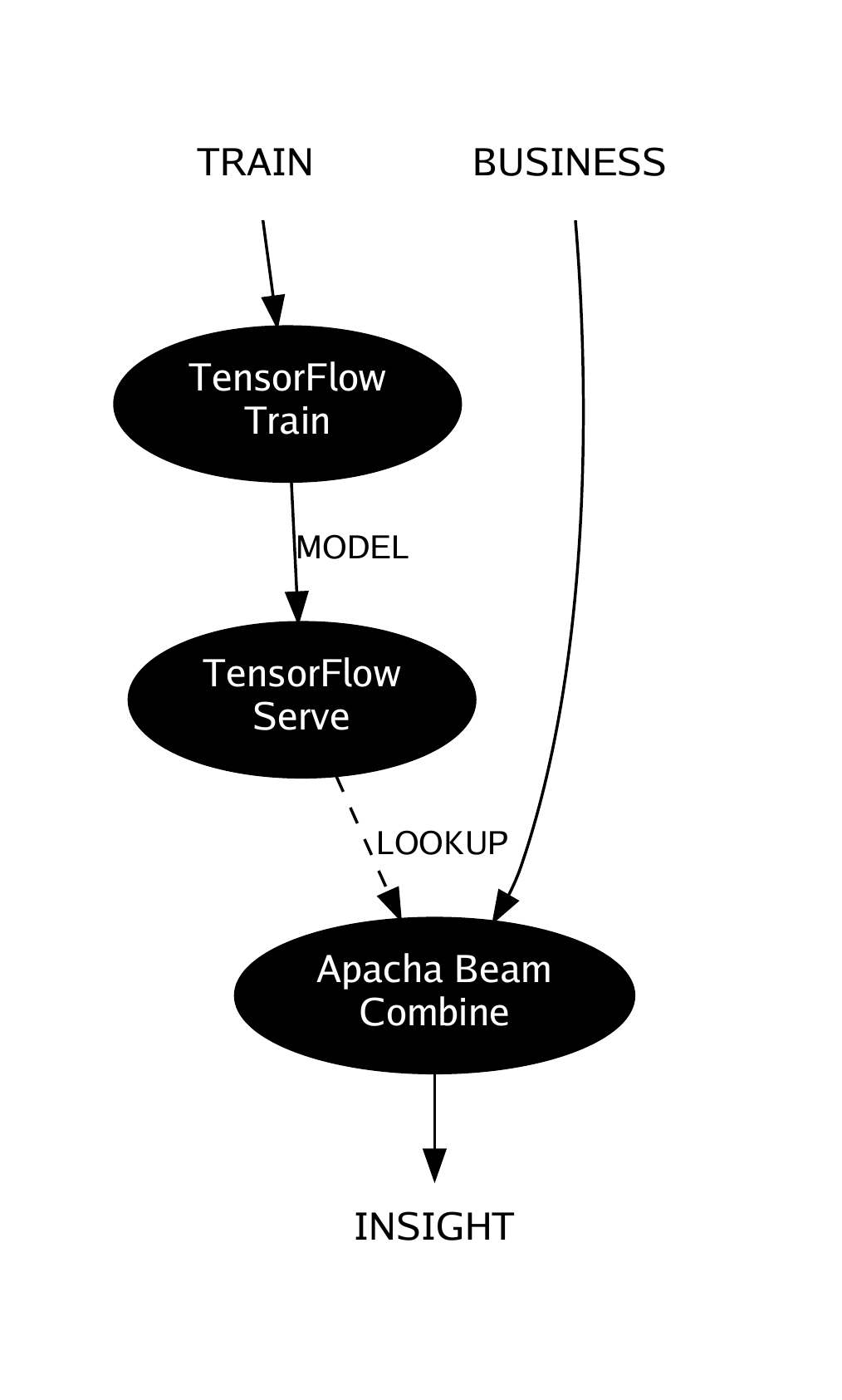}
\centering
\caption{An example workflow which uses different technologies.
Solid edges represent file resources. The dashed edge represents a service resource.}
\end{figure}

The following example captures all semantic aspects of a modern ML/ETL pipeline:

\begin{itemize}
	\item \textbf{INPUTS:} The pipeline expects two input resources from its caller:
	a training table, called TRAIN, and a table of business data, called BUSINESS.

	\item \textbf{STEP 1:} Table TRAIN is used as input to a Machine Learning procedure, e.g. TensorFlow training,
	to produce a new table, we call MODEL. This is a batch job: It processes an input file
	into an output file.

	\item \textbf{STEP 2:} Table MODEL is then used as input to bring up a Machine Model Server,
	e.g. TensorFlow Serve. The server loads the trained model in memory, and
	starts a model-serving API service at a network location, we call SERVICE.

	\item \textbf{STEP 3:} Table BUSINESS together with service SERVICE are used as input to
	an application-specific MapReduce job, which annotates every record in BUSINESS
	with some insight from SERVICE and outputs the result as table INSIGHT.

	\item \textbf{OUTPUT:} Table INSIGHT is the result of the pipeline.
\end{itemize}

A few things are notable here:

\begin{itemize}
	\item[(N1)] The pipelines inputs, intermediate results and outputs are either files or services,
	which we call collectively \textit{resources}

	\item[(N2)] The pipeline program describes a dependency graph between the resources:
	MODEL depends on TRAIN, SERVICE depends on MODEL, and
	INSIGHT depends on BUSINESS and MODEL

	\item[(N3)] The outputs of pipeline steps (be it content of files produced, or behavior of services rendered)
	depend deterministically on their inputs
\end{itemize}

To summarize, this view of a data-processing pipeline captures
resource dependencies and resource semantics (files or services), while treating
computations as black-box deterministic procedures (provided by containers, in practice).

This coarse container/resource-level view of a pipeline suffices to
automate pipeline execution optimally and achieve significant compute and space
efficiencies in common day-to-day operations.

Let us illustrate this with two examples:

\begin{itemize}
\item \textbf{Example 1:} Suppose, after execution, the pipeline completes steps 1 and 2,
then fails during step 3 due to hardware dysfunction.

Due to the determinism (N3) of pipeline steps, it is possible to
cache the file results of intermediate computations, in this case table MODEL,
so they can be reused.

When the pipeline is restarted after its failure, the caching mechanism
would enable it to skip step 1 (a costly training computation) and proceed
directly to restarting the service in step 2 (which takes negligible time)
and then renewing the interrupted computations in step 3.

\item \textbf{Example 2:} In another example, suppose the pipeline is executed successfully
with inputs BUSINESS1 and TRAIN. On the next day, the user 
executes the same pipeline with inputs BUSINESS2 and TRAIN,
due to updates in the business table.

The change in the BUSINESS table does not affect the computations in
step 1 and 2 of the pipeline. Therefore just as in the previous example,
an optimal pipeline would skip these steps and proceed to step 3.
\end{itemize}

%% file: related.tex
\section{Related work: pipeline taxonomy}

Here we position the pipeline technology proposed in this paper against
related technologies in the OSS ecosystem.

For the sake of our comparison, we identify two types of pipeline/workflow
technologies: \textit{task-driven} and \textit{data-driven}. Additionally, data-driven pipelines
are subdivided into \textit{coarse-grain} and \textit{fine-grain} types.

Task-driven pipeline technologies target the execution of a set of 
user tasks, each provided by an executable technology (e.g. binary or container),
according to a dependency graph order. A task executes only after
its dependencies have finished successfully. Task-driven pipelines 
provide simple (usually per-task) facilities for recovering from failure conditions,
like restart rules. In general, task-driven pipelines are not aware of the flow of data
(or services) provided by earlier tasks to later ones.

Data-driven pipeline technologies aim to define and perform
reproducible transformations of a set of input data. The input 
is usually consumed either from structured files (representing things like tables or graphs, e.g.)
located on a cluster file-system, or databases available as services. The outputs are produced in
a similar fashion. Data transformations are specified in the form
of a directed acyclic dataflow graph, comprising data transformations
at the vertices.

The granularity at which the vertex transformations are programmed by the pipeline developer
determines whether a data-driven pipeline is fine-grain or coarse-grain. 

In fine-grain technologies, transformations manipulate data
down to the level of arithmetic and data-structure manipulations of individual
records and their entries. Such transformations are necessarily specified in a
general programming language, where records are typically represented by language values
(like structures, classes, arrays, etc).

In coarse-grain technologies, transformations manipulate batches of data,
which usually represent high-level concepts like tables, graphs, machine learning models, etc.
Such transformations are specified by referring to pre-packaged applications which
perform the transformations when executed.
Coarse-grain transformations are typically programmed using declarative
configuration languages, which aim at describing the dataflow graph and how
to invoke the transformations at its vertices.

Data-driven technologies (fine- or coarse-grain) are aware of the types of
data that flow along edges, as well as
the semantics of the transformations at the vertices (e.g. how they modify the data types).
This allows them to implement efficient recovery from
distributed failures without loss of progress, and more generally
efficient caching everywhere.

The pipeline technology proposed here belongs to the bucket
of data-driven, coarse-grain pipelines which is largely unoccupied
in the OSS space.

%
\begin{table}[]
\begin{tabular}{lll}
 & \textbf{Task-driven} & \textbf{Data-driven} \\ \cline{2-3} 
\multicolumn{1}{l|}{\textbf{Fine-grain}} & \multicolumn{1}{l|}{} & \multicolumn{1}{l|}{\begin{tabular}[c]{@{}l@{}}MapReduce, Spark,\\ Beam, Storm, Heron,\\ Dask\end{tabular}} \\ \cline{2-3} 
\multicolumn{1}{l|}{\textbf{Coarse-grain}} & \multicolumn{1}{l|}{\begin{tabular}[c]{@{}l@{}}Airflow, Argo,\\ Brigade\end{tabular}} & \multicolumn{1}{l|}{Reflow, Dagster, this paper} \\ \cline{2-3} 
\end{tabular}
\end{table}
%

For comparison:

Apache Airflow~\cite{apache_airflow} is task-driven with a configuration-as-code (Python) interface.
Argo is task-driven with a declarative YAML configuration interface.
Brigade is task-driven with an imperative, event-based programming interface.

MapReduce implementations, like Gleam, are fine-grain data-driven.

Apache Spark~\cite{apache_spark} is a fine-grain data-driven pipeline with interfaces in Scala, Java, Python.

Apache Beam~\cite{apache_beam} is a fine-grain data-driven pipeline with interfaces in Go, Java and Python.

Apache Storm~\cite{apache_storm} is a fine-grain data-driven programmable pipeline for distributed
real-time/streaming computations in Java.

Apache Heron~\cite{apache_heron} is a real-time, distributed, fault-tolerant stream processing engine from Twitter;
it is fine-grain data-driven and a direct successor to Apache Storm.

Another interesting take on fine-grain data-driven technologies has emerged in the
Python community, where distributed pipeline programming has been disguised 
in the form of “array” object manipulations. The Dask~\cite{dask} project is one such example,
where operations over placeholder NumPy arrays or Pandas collections is distributed
transparently. Another example is the low-level PyTorch~\cite{pytorch} package
for distributed algorithms~\cite{pytorch_dist}, which allows for passing PyTorch tensors across workers.

Dagster~\cite{dagster} is a course-grain data-driven pipeline with a front-end in Python.
It differs from Koji in that its architecture is not adequate for 
supporting service-based resources (and their automation, caching and garbage-collection).
Aside from its front-end language choice, Dagster is very similar to Reflow, described next.

Reflow~\cite{reflow} is the only coarse-grain data-driven pipeline we found. Reflow 
was designed with a specific use in mind, which makes it fall short of
being applicable in a more general industry setting. Reflow does not
seem to manage data flow across service dependencies, as far as we can tell.
It is AWS-specific, rather than platform agnostic. It is based on a 
new DSL, making it hard to inter-operate with industry practices like configuration-as-code.

Outside of the OSS space, one can find a variety of emerging closed-source or
domain-specific monolithic solutions to coarse-grain data-driven workflows.
One such example are the bio-informatics products of LifeBit~\cite{lifebit}.

%% file: semantics.tex
\section{Semantics}

In this section, we discuss the proposed semantics.

\subsection{Representation}

\subsubsection{Dataflow topology}

A data-processing \textit{pipeline} is represented as a directed acyclic graph,
whose vertices and edges are called \textit{steps} and \textit{dependencies}, respectively.

\begin{itemize}
\item Every pipeline vertex (i.e. step) has an associated set of named \textit{input slots} and
a set of named \textit{output slots}.

\item Every directed pipeline edge (a dependency) is associated with
(1) an output slot at its source vertex, and (2) an input slot at its sink vertex.
\end{itemize}

Output slots can have multiple outbound edges, reflecting that the output
of a step can be used by multiple dependent steps. Input slots, on the other hand,
must have a unique inbound edge, reflecting that a step input argument 
is fulfilled by a single upstream source.

\subsubsection{Steps and transformations}

In addition to their graph structure, steps and dependencies are associated with
computational meaning.

Each dependency (i.e. directed graph edge) is associated with a \textit{resource}, which is
\textit{provided} by the source step and \textit{consumed} by the sink step (of the dependency edge).

Resources are analogous to types in programming languages:
They provide a “compile-time” description of the data processed
by the pipeline at execution time.

Pipeline resource descriptions capture both the data semantics (file or service)
as well as the data syntax (file format or service protocol).

Each step (i.e. graph vertex) is associated with a (description of a) \textit{transform}.
A transform is a computational procedure which, at execution time, 
consumes a set of input resource instances and
produces a set of output resource instances, whose names and resource
types are as indicated by the inbound and outbound edges of the pipeline step.

There are two distinguished transform (i.e. vertex) types,
called \textit{argument} and \textit{return} transforms,
which are used to designate the inputs and outputs of the pipeline itself.
Argument transforms have no input dependencies and a single output dependency.
Return transforms have a single input dependency and no output dependencies.

Steps which are not based on argument or return transforms are called \textit{intermediate}.


\subsection{Execution model}

When a pipeline is executed by a \textit{caller}
(either a human operator or through programmatic control),
a pipeline controller is allocated to dynamically manage the execution of the pipeline
towards the goal of delivering the pipeline's return resources to the caller.

The key technical challenge in designing the pipeline control logic is
to devise a generic algorithm which is robust against process failures,
while also accommodating for the semantic differences between
file and service resources:

\begin{itemize}
	\item	File resources are considered available \textit{after} the successful termination of
		the transformation process that produces them,

	\item	Service resources are considered available \textit{during} the execution of
		the transformation process that produces them.
\end{itemize}

\subsubsection{Control algorithm}

The pipeline execution algorithm, performed by the \textit{pipeline controller},
associates two state variables with each dependency (edge) in the pipeline graph.

\begin{itemize}
	\item	A variable that assumes one of the states “available” or “not available”,
	indicates whether the underlying resource (file or service) is currently available.
	This variable is written by the supervisor (see below) of the step producing the dependency,
	and read by the supervisor of the step consuming the dependency.

	\item A variable that assumes one of the states “needed” or “non needed”,
	indicating whether the underlying resource (file or service) is currently needed.
	This variable is written by the supervisor of the step consuming the dependency,
	and read by the supervisor of the step producing the dependency.
\end{itemize}

On execution, the pipeline controller proceeds as follows:

\begin{enumerate}
	\item Mark the state of each input dependency to a return step as “needed”.
	These dependencies will remain “needed” until the pipeline is terminated by the caller.

	\item For each intermediate step in the pipeline graph,
	create a step \textit{supervisor}, running in a dedicated process (or co-routine).
\end{enumerate}

Every step supervisor comprises two independent sub-processes:
a \textit{driver} loop and a \textit{(process) collector} loop.

The driver loop is responsible for “sensing” when the outputs of
the supervised step are needed dynamically (by dependent steps) and arranging
for making them available.

\begin{enumerate}
	\item\label{itm:controller:loop}	Repeat:
	\begin{enumerate}
		\item	If the step has no output dependencies
					which are “needed” and “not available”, then goto (\ref{itm:controller:loop}).
		\item	Otherwise:
		\begin{enumerate}
			\item	Mark all input dependencies of the step as “needed”
			\item	Wait until all input dependencies become “available”
			\item	Start the container process associated with this step
			\item	After starting the process, mark all service output dependencies of this step as “available”
			\item	Wait until the process terminates:
			\begin{enumerate}
				\item	If the termination state is “succeeded”  (implying that all output files have been produced), then:
						Mark all file output dependencies of this step as “available”.
						(Since output files are cached, these dependencies will remain in state “available”.)
						Mark all input dependencies of this step as “not needed”.
						Goto (\ref{itm:controller:loop}).
				\item	If the termination state is “failed”, then:
						Goto (\ref{itm:controller:loop}).
				\item	If the termination state is “killed” (by the collector loop), then:
						Mark all service output dependencies of this step as “unavailable”.
						Mark all input dependencies of this step as “not needed”.
						Goto (\ref{itm:controller:loop}).
			\end{enumerate}
		\end{enumerate}
	\end{enumerate}
\end{enumerate}

The (process) collector loop is responsible for sensing when the outputs of
the supervised step (there is a collector for each step in the pipeline)
are not needed any longer and arranging to garbage-collect
its process.

\begin{enumerate}
	\item\label{itm:collector:loop}	Repeat:
	\begin{enumerate}
		\item	If the step process is running and all of the following conditions hold, then kill the process:
		\begin{enumerate}
			\item	All file output dependencies of the step are
					either “needed” and “available” or “not needed”, and
			\item	All service output dependencies of the step are “not needed”.
		\end{enumerate}
		\item Goto (\ref{itm:collector:loop}).
	\end{enumerate}
\end{enumerate}


\subsection{Pure functions and causal hashing of content}

Most data-processing pipelines in industry are required, by design,
to have reproducible and deterministic outcomes. This includes
workflows such as Machine Learning, banking and finance, canary-ing,
software build systems, continuous delivery and integration, and so on.

In all reproducible pipelines, by definition, step transformations are \textit{pure}:
The outcomes (files output or services provided) obtained from executing
pure transformations are entirely determined by the inputs provided to them
and the identity (i.e. program description) of the transformation.

By contrast, non-reproducible pipelines are ones where transformation outcomes
might additionally be affected by:

\begin{itemize}
	\item runtime information (like the value of the wall clock or the temperature of the CPU), or 
	\item interactions with an external stateful entity (like disk, a persistent store service, 
	or outside Internet services, for instance).
\end{itemize}

\subsubsection{Caching}

In the case of reproducible pipelines (comprising pure transformations), 
pipeline execution can benefit from dramatic efficiency gains (in computation,
communication and space), using a simple technique we call \textit{causal caching}.

The results of pipeline steps which are based on purely deterministic
transformations can be cached to obtain significant efficiency gains
in the following situations:

\begin{enumerate}
	\item Avoiding duplicate computations when restarting partially-executed pipelines,
	for instance, after a hardware failure;
	\item Multiple executions of the same pipeline,
	perhaps by different users concurrently or at different times;
	\item Executions of pipelines that have similar structure, for instance, as is the
	case with re-evaluating the results of multiple incremental changes of the same base pipeline
	during development iterations.
\end{enumerate}

The caching algorithm assigns a number, called a \textit{causal hash}, to
every edge of the computation graph of a pipeline. These hash numbers
are used as keys in a cluster-wide caching file system.

To serve their purpose of cache keys for the outputs of pipeline steps,
causal hashes have to meet two criteria:

\begin{enumerate}
	\item[(C1)]\label{itm:c1} A causal hash has to have the properties of a content hash:
	If the causal hashes of two resources are identical, then the resources
	must be identical.
	\item[(C2)]\label{itm:c2} A causal hash has to be computable before the resource it describes
	has been computed, by executing the step transform that produces it.
\end{enumerate}

To meet these criteria, we define causal hashes in the following manner:

\begin{enumerate}
	\item The causal hashes of the resources passed as inputs to the pipeline
	are to be provided by the caller of the pipeline. Criteria (C2) does not
	apply to input resource, thus any choice of a content hashing algorithm,
	like using an MD5 message digest or a semantic hash, suffices.

	\item All other pipeline edges correspond to resources output by a 
	transformation step. In this case, the causal hash of the resource
	is defined recursively, as the message digest (e.g. using SHA-1) of the following
	meta information:
	\begin{enumerate}
		\item The pairs of name and causal hash for all inputs to the step transformation,
		\item The identity (or program description) of the transformation,
		\item The name of the transformation output associated with the edge.
	\end{enumerate}
\end{enumerate}

Note that while only file resources can be cached (on a cluster file system),
service resources can also benefit from caching. For instance,
a service resource in the middle of a large pipeline, can be made available
if the file resources it depends on have been cached from a prior execution.


\subsubsection{Locking and synchronization}

Pipeline semantics make it possible to execute multiple racing
pipelines in the same cluster, while ensuring they
utilize computational resource optimally.

Two different pipeline graphs can entail similar transformations
in the sense of a common computational subgraph,
appearing in both pipelines.

This situation occurs, for instance, as a developer iterates
over pipeline designs incrementally, producing many similar designs.

A causal cache (as described earlier) shared between 
concurrent pipelines enables one pipeline
to reuse the computed output of an identical step,
that was already computed by the other pipeline.

We accomplish cache sharing across any number of 
concurrently executing pipelines by means of
per-causal-hash cluster-wide locking.

In particular, the controller algorithm for executing a pipeline transformation step
is augmented as follows:

\begin{enumerate}
	\item Compute the causal hashes, H, of the step outputs
	\item Obtain a cluster-wide lock on H
	\item Check if the output resources (files) have already been cached
	in a designated caching file system:
	\begin{enumerate}
		\item If so, then release the lock on H and reuse the cached resources.
		\item Otherwise, execute the step transformation, cache its outputs, 
		release the lock on H and return the output resources.
	\end{enumerate}
\end{enumerate}


\subsubsection{Composability}

The reader will note that a pipeline can be viewed as a transform:
It has a set of named inputs (the arguments), a set of named outputs (the return values)
and a description of an executable procedure (the graph).

Consequently, one pipeline can be invoked as a step transformation in another.

This generic and modular flexibility enables developers to create pipeline templates
for common workflows, like a canary-ing workflow or an ML topology, and
reuse those templates as building blocks in multiple applications.

%% file: arch.tex
\section{Architecture}

Our goal here is to describe an architecture for a data-processing pipeline system, and
outline an implementation strategy that works well with available OSS software.

We focus on an approach that uses Kubernetes as underlying infrastructure,
due to its ubiquitous deployments in commercial clouds.

The pipeline execution logic itself is implemented as a Go library,
which can execute a pipeline given runtime access to a Kubernetes
cluster and a user pipeline specification.
Pipeline executions can be invoked through standard integration points:
(a) using a command-line tool by passing a pipeline description file,
(b) using a Kubernetes controller (via CRD), or (c) from any programming
language by sending pipeline configurations for execution to the controller
interface in (b).

The approach (c) is sometimes called configuration-as-code and is a common
practice. For instance, TensorFlow and PyTorch are Python front-ends for
expressing tensor pipelines. Using general imperative languages to express
pipelines has proven suboptimal for various reasons. For one, pipelines
(in the sense of DAG data flows) correspond to immutable functional
semantics (not imperative mutable ones). Furthermore, configuration-as-code
libraries have not been able to deliver type-safety at compile-time.
To solve for both of these problems, we have designed a general functional
language, called Ko~\cite{ko}, which allows for concise type-safe functional-style
expression of pipelines. 


\subsection{Type checks before execution}

The pipeline specification schema allows the user to
optionally specify more detailed “type” information about
the resources input to or output by each transformation
in a dataflow program.

For file resources, this type information can describe the
underlying file and its data at various levels of precision.
It could specify a file format (e.g. CSV or JSON),
an encoding (e.g. UTF8) and a data schema (e.g. provided
as a reference to a Protocol Buffer or XML schema definition).

For service resources, analogously, the user can
optionally describe the service type in to a varying
level of detail: transport layer (e.g. HTTP over TCP),
security layer (e.g. TLS), RPC semantics (e.g. GRPC),
and protocol definition (e.g. a reference to a Protocol
Buffer or an OpenAPI specification).

When such typing information is provided, the pipeline 
controller is able to check the user's dataflow programs
for type-safety, before it commits to a lengthy execution,
as is often the case.


\subsection{Resource management and plumbing}

At the programming level, the user directly connects
the outputs of one transformation to the inputs of another.

At runtime, however, these intermediate
resources — be it files or services — need to managed.

\subsubsection{Files}

For intermediate file resources, generally, the pipeline controller will
determine the placement of files on a cluster volume and will connect
these volumes as necessary to containers requiring access to the files.

For instance, assume transformation A has an output that is connected to an
input of transformation B. At runtime, the controller will choose a volume for
placing the file produced by A and consumed by B.
It will attach this volume to the container for A during its execution,
and then it will attach the volume (now containing the produced file)
to the container for B. Plumbing details such as passing execution flags
to containers are handled as well.

Of course, this is a basic example. The file management logic can
be extended with hooks to fulfill various optimization and policy needs, such as:

\begin{itemize}
	\item Plugging third-party file placement algorithms that optimize physical placement locality, or
	\item Placing non-cacheable resources on memory-backed volumes,
\end{itemize}

The file management layer also contains the causal-hash-based caching
of files (described in the previous section).

\subsubsection{Services}

For intermediate services resources — provided from one transformation to the
next — the pipeline controller handles plumbing details transparently, as
well. Generally, it takes care of creating DNS records for services, and
coordinating container server addresses and flag-passing details.

As with files, services between a server and a client transformation, can be customized
via hooks to address load-balancing, re-routing, authentication, and other such concerns.


\subsection{Transformation backends}

A transformation is, generally, any process execution
within the cluster, which accepts files or services as inputs,
and produces files or provides services as output.

From a technology point of view, a transform can be:

\begin{enumerate}
	\item The execution of a container,
	\item The execution of a custom controller, known as a Kubernetes CRD.
	For instance, the \texttt{kubeflow} controller is used to start TensorFlow jobs against
	a running TesnorFlow cluster (within Kubernetes),
	\item More generally, the execution of any programming code that
	orchestrates the processing of input resources into output resources.
\end{enumerate}

To accommodate such varying needs, Koji provides a simple
mechanism for defining new types of transforms as needed.

From a system architecture perspective, a transform comprises two parts:

\begin{enumerate}
	\item A schema for the declarative configuration that the user provides to instantiate
	transforms of this type, and
	\item A backend implementation which performs the execution,
	given a configuration structure (and access to the pipeline and cluster APIs).
\end{enumerate}

Optionally, such backends can install dependent technologies during an
initialization phase. For instance, a backend for executing Apache Spark jobs might
opt to include an installation procedure for Apache Spark, if it is not present
on the cluster.

This paper uses container execution as the running example throughout the sections
on semantics and specification. In practice, most legacy/existing OSS technologies
will require a dedicated backend, due the large variety of execution and installation
semantics. 

Fortunately, writing such backends is a short one-time effort. One can envision 
amassing a collection of backends for common technologies like Apache Spark, 
Apache Beam, TensorFlow, R, and so on.

Each such technology will define a dedicated configuration structure,
akin to \texttt{Container} (in the specification section), which captures the parameters
needed to perform a transform execution. We believe that such a simple-to-use
declarative library of transforms backed by OSS technologies provides
standardized assembly-level blocks for expressing business flows, in general.


\subsection{Transform job scheduling}

The pipeline controller orchestrates the execution of
transforms in their dependency order: A transformation step
is ready to execute only when the resources it depends on
become available.

Beyond this semantic constraint on execution order,
transform jobs can be scheduled to meet additional
optimization criteria like throttling or locality of placement.

Optimized job scheduling (and placement) is generally provided
by various out-of-the-box products, like Medea~\cite{medea}, for instance.

Since job scheduling considerations are orthogonal to the
pipeline execution order, our architecture makes it easy to
plug in any scheduling algorithm available out-of-the-box.

This is accomplished by implementing a short function
which the pipeline controller uses to communicate with the scheduler when needed.
Transform steps can be annotated (in the pipeline's declarative
configuration) with tags that determine their scheduling affinities
which would be communicated to the scheduler.


\subsection{Recursion}

Pipelines can be executed from multiple places in the software stack, e.g.

\begin{enumerate}
	\item Using a command-line tool, which consumes a pipeline configuration file (YAML or Protocol Buffers),
	\item Using a Kubernetes CRD whose configuration schema understands the pipeline schema shown here,
	\item From any program running in the cluster, using a client library, also
	by providing a pipeline configuration as an execution argument.
\end{enumerate}

In particular, for instance as implied by the last method, a program that runs as a 
transform in one pipeline can — as part of its internal logic — execute another
pipeline.

In the presence of such recursive pipeline invocation, all data consistency
and caching guarantees remain in effect, due the powerful nature of causal hashes.
This enables developers to build complex recursive pipelines,
such as those required by Deep Learning and Reinforcement Learning methodologies.

Due to the ability to execute pipelines recursively and
the modular declarative approach to defining pipelines as configurations,
our pipeline system can directly be reused as the backend of
a DSL for programming data-processing logic at the cluster-level.

We have made initial strides in this direction with the design and implementation
of the Ko programming language. We defer this extension to a follow up paper.

%% file: conclusion.tex
\section{Conclusion}

This paper has two main contributions. 
First, we make the observation that virtually
all large-scale, reproducible data-processing pipelines
follow a common pattern, when viewed at the right level of abstraction. 

In particular, at a semantic level, said pipelines can be viewed as
dependency-based build tools for data, akin to code build tools for software.
Within this context, however, pipelines differ from build tools for code in that
the resources being depended on can be files as well as short-lived services.

Our second contribution is to cast these two types of resources —
which have very different runtime semantics — into a unified framework,
where either can be viewed merely as a simple "resource dependency"
from the point of view of the user.

To make this possible, we introduce Causal Hashing which
is a method for generating content hashes for both files and services.
Causal Hashing is thus a generalization of content hashing,
which can assign unique content IDs to complex temporal objects
(like services).

Causal Hashing unlocks the complete automation of a myriad of
tasks, such as caching, conflict resolution, version tracking, incremental builds
and much more.

%% file: spec.tex
\section{Specification}

\subsection{Specification methodology}

Programmable technologies, in general, expose the user-programmable
functionality in one of two ways. Either by using a (general or domain-specific)
programming language, or using a typed declarative schema (captured by
standard technologies like XML, YAML, Thrift or Protocol Buffers, for instance)
for expressing program configurations.

For instance, Apache Spark and Apache Storm are programmable through Java.
Gleam, a MapReduce reduce implementation in Go, is programmable through Go.
On the other hand, TensorFlow and Argo express their programs in the form
of computational DAGs captured by Protocol Buffers and YAML, respectively.

The use of typed data structures, in the form of Protocol Buffers, for defining
programmable software has been a wide-spread practice within Google,
for many years now. 

This declarative/configuration approach has a few advantages.

The configuration schema for any particular
technology acts as an “assembly language” for that technology 
and provides a formal decoupling between programmable semantics and
any particular implementation. Furthermore, declarative configurations
being data succumb to static analyses (for validity, security or policy, e.g.)
prior to execution, which is not the case with DSL-based interfaces.

Configuration schema are language-agnostic, as they can be generated
from any general programming language: a practice widely used
and known as configuration-as-code.

The declarative schema-based approach is gaining momentum in the OSS space as well,
as witnessed for instance by projects like ONNX. ONNX defines 
a platform-independent programming schema for describing ML models
in the form of an extensible Protocol Buffer. ONNX aims to be viewed
as a standard, to be implemented by various backends.

In this spirit, we believe that the correct interface for defining
a general-purpose data-processing pipeline is the typed declarative one.
We use Protocol Buffers as they provide a time-tested extension mechanism
for the definition of backward- and forward-compatible data schemas.
But it should be understood that interoperability with other standards like
OpenAPI and YAML is a given, using standard tooling.


\subsection{Pipeline}

A \textit{pipeline} is a directed acyclic graph whose vertices,
called pipeline \textit{steps}, represent data-processing tasks and
whose edges represent (file or service) dependencies between
pairs of steps.

At the highest level, a pipeline is captured by message \texttt{Pipeline}, shown below.

\begin{lstlisting}
message Pipeline {
	repeated Step steps = 1;
}
\end{lstlisting}

A pipeline is an executable application, which will (1) consume some
inputs from its cluster environment (e.g. files and directories from a volume,
or a stream of data from a micro-service API), (2) process these inputs through
a chain of transformation steps, and (3) deliver some outputs (which could be
data or services).

\subsection{Steps}

A pipeline \textit{step} is the generic building-block of a pipeline application.
Steps are used to describe the inputs, intermediate transformations, and
outputs of a pipeline application.

A step is captured by message \texttt{Step} below:

\begin{lstlisting}
message Step {
	required string label = 1;
	repeated StepInput inputs = 2;
	required Transform transform = 3;
}
\end{lstlisting}

Each step is identified by a unique string \textit{label}, which distinguishes
it from other steps in the pipeline. This is captured by field \texttt{label}.

The step definition specifies the transformation being
performed by the step, as well as the sources for the transformation's inputs
relative to the pipeline.

The step's transformation is captured by field \texttt{transform}.
Transformations are self-contained descriptions of data processing logic
(described in more detail later).

Each transformation declares a list of named and typed inputs
(which can be viewed akin to functional arguments), as well as 
a list of named and typed outputs (which can be viewed akin to
functional return values).

Field \texttt{inputs} describes the source of each named input, expected by
the step's transformation. Each named input is matched with
another step in the pipeline, called a \textit{provider}, and a specific
named output at the provider step.

This matching between inputs and provider steps is captured in 
message \texttt{StepInput} below:

\begin{lstlisting}
message StepInput {
	required string name = 1;
	required string provider_step_label = 2;
	required string provider_output_name = 3;
}
\end{lstlisting}


\subsection{Transform}

A \textit{transform} is a self-contained, reusable description of a data-processing
computation, based on containerized technology.

Akin to a function (in a programming language), a transform comprises:
(1) a set of named and typed inputs,
(2) a set of named and typed outputs, and
(3) an implementation, which describes how to perform the transform using containers.

Transforms are described by message \texttt{Transform} below.

\begin{lstlisting}
message Transform {
	repeated TransformInput inputs = 1;
	repeated TransformOutput outputs = 2;
	required TransformLogic logic = 3;
}
\end{lstlisting}

\subsubsection{Transform inputs and outputs}

Transform inputs and outputs are captured by
messages \texttt{TransformInput} and \texttt{TransformOutput} below.

\begin{lstlisting}
message TransformInput {
	string name = 1;
	Resource resource = 10;
}
message TransformOutput {
	string name = 1;
	Resource resource = 10;
}
\end{lstlisting}

The inputs (and outputs) of a transformation are identified by unique names.

These names serve the purpose to decouple the pipeline wiring definitions
(captured in messages \texttt{Step} and \texttt{StepInput}) from the implementation
detail of how inputs are passed to the container technology backing
a transform (captured within message \texttt{TransformLogic}).

Each input (and output) is associated with a \textit{resource} type,
which is captured in field \texttt{resource}.


\subsection{Resources}

A \textit{resource} is something that a transform consumes
as its input or produces as its output.

The type of a resource is defined using message \texttt{Resource} below.
A \texttt{Resource} should have exactly one of its fields, \texttt{file} or \texttt{service}, set.

\begin{lstlisting}
message Resource {
	optional FileResource file = 1;
	optional ServiceResource service = 2;
}
\end{lstlisting}

Resource types capture the “temporal” nature of a
resource (e.g. file vs service), as well as its “spacial” nature
(e.g. specific file format or specific service protocol).

Resource type information is used in two ways by the pipeline controller:

\begin{enumerate}
\item To verify the correctness of the step-to-step pipeline stitching in an
application-meaningful way. Specifically, the pipeline compilation process
will verify that the resource output by one step \textit{fulfills} the resource type
expected as input by a downstream dependent step.

\item To inform garbage-collection of steps that provide services as their output.
Specifically, if a step provides a service resource as its output,
the pipeline controller will garbage-collect the step (e.g. kill its underlying 
container process) as soon as all dependent steps have completed their
tasks. In contrast, a step which provides file resources will be garbage-collected
only after it terminates successfully on its own.
\end{enumerate}

\subsubsection{File resources}

A file resource type is described using message \texttt{FileResource}:

\begin{lstlisting}
message FileResource {
	required bool directory = 1;
	optional string encoding = 2;
	optional string format = 3;
}
\end{lstlisting}

The file type specifies whether the resource is a file or directory,
and associates with it an optional encoding and an optional format identifier.

Encoding and format identifiers are used during pipeline compilation
to verifying that the output resource type of a provider step
fulfills the input resource type of a consumer step. In this context,
if provided, the encoding and format identifiers are treated as opaque strings
and are checked for exact match.

\subsubsection{Service resources}

A service resource type is described using message \texttt{ServiceResource}:

\begin{lstlisting}
message ServiceResource {
	optional string protocol = 1;
}
\end{lstlisting}

The service type optionally specifies a protocol identifier.

This identifier is used during pipeline compilation to ensure that
the service provided by one step's output fulfills the service
expectations of a dependent step's input.
In this context if the protocol identifier is given on both sides,
it will be verified for an exact match.

Protocol identifiers should follow a meaningful convention which,
at minimum, determines the service technology (e.g. GRPC vs OpenAPI)
and the service itself (e.g. using the Protocol Buffer fully-qualified name
of the service definition). For example, 

\begin{lstlisting}
openapi://org.proto.path.to.Service"
\end{lstlisting}


\subsection{Transform logic}

The \textit{logic} of a transform is a description (akin to a function implementation)
of what a transform does and how it does it.

Transform logic is described by message \texttt{TransformLogic} shown below.

\begin{lstlisting}
message TransformLogic {
	optional ArgumentLogic arg = 100;
	optional ReturnLogic return = 200;
	optional ContainerLogic container = 300;
	// additional logics go here, e.g.
	// optional TensorFlowLogic tensor_flow = 301;
	// optional ApacheKafkaLogic apache_kafka = 302;
	// etc.
}
\end{lstlisting}

Message \texttt{TransformLogic} consists of a collection of mutually-exclusive logic types
captured by the message fields, of which exactly one must be set.
Each logic type is implemented as a “plug-in” in the pipeline controller
and additional logics can be added, as described in the section on architecture.

\subsection{Pipeline arguments}

Pipelines, like regular functions, can have arguments whose values are supplied at execution time.
Unlike function argument values (which are arithmetic numbers and data structures),
pipeline argument values are resource (file or service) instances.

The dedicated transform logic, called \textit{argument}, is used to declare pipeline arguments.

From a pipeline graph point of view, steps based on argument transforms
are vertices that have no input edges and a single output edge, representing
the resource supplied to the argument when the pipeline was executed.

Message \texttt{ArgumentLogic}, shown below, describes a pipeline argument.

\begin{lstlisting}
message ArgumentLogic {
	required string name = 1;
	required FileResource resource = 2;
}
\end{lstlisting}

Field \texttt{name} specifies the name of the pipeline argument.
Field \texttt{resource} specifies the type of file or service resource expected as argument value.

Argument steps have one output in the pipeline graph,
whose resource type is that provided by field \texttt{resource}.

\subsection{Pipeline return values}

Pipelines, like regular functions, can return values to the caller environment.
In the case of pipelines, the returned values are resource (file or service) instances.

The dedicated transform logic, called \textit{return}, is used to declare pipeline return values.

From a pipeline graph point of view, steps based on a return transform
are vertices that have a single input edge, representing the resource to be returned
to the pipeline caller, and no output edges.

Message \texttt{ReturnLogic}, shown below, describes a pipeline return value.

\begin{lstlisting}
message ReturnLogic {
	required string name = 1;
	required Resource resource = 2;
}
\end{lstlisting}

Field \texttt{name} specifies the name of the return value.
Field \texttt{resource} specifies the type of file or service resource returned.

Return steps have one input in the pipeline graph,
whose resource type is that provided by field \texttt{resource}.


\subsection{Container-backed transforms}

A \textit{container logic} describes a pipeline transform backed by a container.

Container logics are captured by message \texttt{ContainerLogic} shown below.

\begin{lstlisting}
message ContainerLogic {
	required string image = 10;
	repeated ContainerInput inputs = 20;
	repeated ContainerOutput outputs = 21;
	repeated ContainerFlag flags = 22;
	repeated ContainerEnv env = 23;
}
\end{lstlisting}

The container logic specification captures:

\begin{enumerate}
\item The identity of the container image, e.g. a Docker image label.
This is  captured by field \texttt{image} of message \texttt{Container}.

\item For every named transform input and output (declared in
fields \texttt{inputs} and \texttt{outputs} of message \texttt{Transform}),
a method for passing the location of the corresponding resource (file or service) to
the container on startup. Methods for passing resource location include flags
and environment variables, as well as different formatting semantics, and are described later.
Input and output passing methods are captured by fields \texttt{inputs} and \texttt{outputs}
of message \texttt{Container}.

\item Any additional startup parameters in the form of flags or environment variables.
These are captured by fields \texttt{flags} and \texttt{env} of message \texttt{Container}.
\end{enumerate}

\subsubsection{Container input and output}

Messages \texttt{ContainerInput} and \texttt{ContainerOutput} associate
every transform input and output, respectively,  with a container
flag and/or environment variable, where the resource locator
is to be passed to the container on startup.

\begin{lstlisting}
message ContainerInput {
	required string name = 1;
	optional string flag = 2;
	optional string env = 3;
	optional ResourceFormat format = 4;
}
message ContainerOutput {
	required string name = 1;
	optional string flag = 2;
	optional string env = 3;
	optional ResourceFormat format = 4;
}
\end{lstlisting}

Both messages have analogous semantics:

Field \texttt{name} matches the corresponding transform input (or output) name,
as declared in message \texttt{Transform} within field \texttt{inputs} (or \texttt{outputs}).

Fields \texttt{flag} and \texttt{env} determine the flag name and environment variable,
respectively, where the resource locator is to be passed.

Field \texttt{format} determines the resource locator formatting convention.

By default, file resource location is passed to the container in the form of
an absolute path relative to the container's local file system.

By default, service resource location is passed to the container using
standard host and port notation.

Alternatives, are provided by message \texttt{ResourceFormat}.

\subsubsection{Container parameters}

Additional container parameters, specified in fields \texttt{flags} and \texttt{env} of message
\texttt{Container}, are defined by message \texttt{ContainerFlag} below.

\begin{lstlisting}
message ContainerFlag {
	required string name = 1;
	optional string value = 2;
}
\end{lstlisting}